# Realization of cavity linewidth narrowing via interacting dark resonances in a tripod-type electromagnetically induced transparency system

Kang Ying,[1] Yueping Niu,[2,∗] Dijun Chen,[1,†] Haiwen Cai,[1] Ronghui Qu,[1] and Shangqing Gong[2,‡]

[1]*Shanghai Institute of Optics and Fine Mechanics,*
*Chinese Academy of Sciences, Shanghai 201800, China*
[2]*Department of Physics, East China University of Science and Technology, Shanghai 200237, China*

compiled: November 22, 2013

Cavity linewidth narrowing via double-dark resonances has been experimentally observed using the $^{87}$Rb Zeeman splitting sublevels. With the steep dispersion led by the interacting dark resonances in the tripod-type electromagnetically induced transparency system, we narrow the cavity linewidth to 250 KHz at room temperature. Furthermore, the position of this ultranarrow cavity linewidth could be tuned in a 60 MHz coupling field detuning range.

## 1. Introduction

Dark resonance is by now a well-known concept in optics and laser spectroscopy. It has attracted considerable attention because of its potential applications in many fields [1–16], such as the resonant enhancement of the refractive index, quantum entanglement, optics image processing etc. The essential feature of dark resonance is the existence of quantum superposition states, which are decoupled from the coherent and dissipative interactions. When more than one coupling fields exist, coherent coupling fields would dress more atomic states and multiple quantum superposition states appear. The coherent interaction involving the multiple quantum superposition states leads to the splitting of dark states, so called double-dark resonances. Double-dark resonances were first studied by Lukin et al. [17] and they predicted the possibility of sharp, high contrast resonances. Later, the interacting dark resonances induced quantum interference phenomenon was experimentally observed in [18] and different schemes of double-dark resonances were explored [19–22].

As a typical effect of dark resonance, electromagnetically induced transparency (EIT) has attracted great attention because of the properties of large dispersion and almost vanishing absorption in such system. Many studies have been done to narrow the optical cavity linewidth via placing an EIT medium in an ordinary cavity, known as intracavity EIT. Large dispersion and vanishing absorption would result in a substantial narrowing of cavity linewidth. Lukin and co-workers first discussed theoretically the effect of intracavity EIT on the properties of optical resonators and predicted the pronounced frequency pulling and cavity linewidth narrowing effect [23]. Then, intracavity experiments were made to narrow the cavity linewidth to about 1 MHz via Lambda-type EIT using hot $^{87}$Rb atoms at 87°C [24, 25] and cold $^{87}$Rb atoms at MOT [26]. Very recently, we have narrowed the cavity linewidth to 1.2 MHz via the V-type EIT using the $^{85}$Rb atoms at room temperature [27]. All the above intracavity studies are based on the single-dark resonance, while in the double-dark resonances case, the interacting dark states can lead to a narrower transparency window and the dispersion in the narrower transparency window is much larger than the one of the single-dark system. Recently, we have theoretically investigated the ultranarrow cavity linewidth in a tripod-type double-dark resonances EIT system [28]. It was predicated that the narrowed cavity linewidth via double-dark resonances could be one order of magnitude narrower than that of the single-dark system and the position of the ultranarrow transmission peak could be controlled by adjusting the detuning of the coupling field. Here, we will report the experimental observation of the cavity linewidth narrowing via double-dark resonances with the tripod-type Zeeman splitting sublevels of $^{87}$Rb inside. The cavity linewidth could be narrowed to 250 KHz at room temperature, which is five times narrower than the single-dark resonance case in the same experimental condition. The ultranarrow linewidth can be realized in a 60 MHz

∗ niuyp@ecust.edu.cn
† djchen@siom.ac.cn
‡ sqgong@ecust.edu.cn

frequency range.

## 2. Tripod-type configuration

The relevant energy levels used in our experimental study is the $D_2$ line of $^{87}$Rb, as Fig. 1 shows. Two strong coupling beams $E_1$ and $E_2$ drive the transition F=2 to F'=1 with detuning $\Delta_1$ and $\Delta_2$, while the weak probe beam $E_p$ is scanned across the level F=1 to F'=1. When we apply a longitudinal magnetic field, each sublevels are shifted by an amount $\mu_B g_F m_F B$, where $\mu_B = 1.4$ MHz/G is the Bohr magnetron, $g_F$ is the Landé factor of the levels, and $B$ is the magnetic field. In order to simplify the complicated Zeeman splitting sublevels, we transform the polarization states of the laser beams $E_p$, $E_1$ and $E_2$ from linear to circular polarization when they enter into the Rb vapor cell using two quarter-wave plates (QWP). Adjusting the optical axis angle of the QWPs appropriately, the probe and coupling beams will be seen in the atoms' frame as being $\sigma^+$ or $\sigma^-$ circularly polarized. The detailed polarization states of the probe beam and two coupling beams are shown in Fig. 2. In this figure, all the beams propagate in the same direction inside the vapor cell and the circular polarizations are defined in this direction. As a result, the $\sigma^+$ polarized probe beam $E_p$ and coupling beam $E_2$ could only motivate the transition $\Delta m_F = +1$, while the $\sigma^-$ polarized coupling beam $E_1$ could only motivate the transition $\Delta m_F = -1$. Thus, a tripod-type configuration is formed using $^{87}$Rb atoms Zeeman splitting sublevels. In such a tripod-type system, the additional transition to the $\Lambda$-type single-dark system by another coupling field causes the occurrence of two distinct dark states. According to our previous study [28], interacting dark states would make one of the cavity transmission peaks one order of magnitude narrower than that of single-dark system and this ultranarrow spectrum can keep in a wide coupling field detuning range.

## 3. Experiment and discussions

A diagram of the experimental setup is shown in Fig. 3. The probe laser, coupling laser 1 and coupling laser 2 are single-mode tunable external cavity diode lasers (ECDL) (New Focus TLB-6900), which have a linewidth of 300 KHz (measured using beat frequency method). The half-wave plate 1 (HWP1) and polarized beam splitter 1 (PB1) are used to attenuate the power of probe laser to below $50\mu$W to avoid the saturated absorption of the $^{87}$Rb atoms and self-focusing effect. About 10% of the coupling lasers power is separated into auxiliary Rb cells for monitoring the frequency of coupling laser 1 and coupling laser 2. The probe beam $E_p$ and coupling beam $E_1$ are brought together by the polarized beam splitter 2 (PB2). The coupling beam $E_2$ is reflected into the Rb vapor cell by the polarized beam splitter 3 (PB3) and a mini-reflective mirror M3 adhered to the edge of the Rb vapor cell. Thus, all the laser beams are co-propagated to minimize the residual Doppler linewidth. The two acoustic optical modulators (AOM) are used to modulate the frequency of two coupling beams. The probe beam and the two coupling beams (not circulating in the cavity) are focused into the cavity by lenses with focal length of 30 cm and their respective beam diameters at the center of the Rb vapor cell are all about $200\mu$m. After passing through the quarter-wave plate 1 (QWP1) and quarter-wave plate 2 (QWP2), both the coupling beams and the probe beam are circular polarization when they enter the 75 mm-AR-coated Rb vapor cell. A solenoidal coil surrounds the Rb vapor cell to supply a longitudinal magnetic. The HWP2 and HWP3 are used to adjust the power of two coupling beams. The mirrors M1 and M2 are used to form a Fabry-Pérot (FP) cavity and the disturbance of the mini-reflective mirrors M3 to the cavity is not significant as the space of M3 is very tiny (diameter less than 2 mm). The reflectivity of cavity mirror M1 is approximately 99.5%. The cavity mirror M2 with a reflectivity of 99.5% is controlled by a piezoelectric (PZT) driver. Both the mirrors M1 and M2 are concave with a 55 cm radius of curvature. After we insert the Rb cell, PB2, PB3, QWP1 and QWP2, the finesse of the cavity (with the Rb atoms off resonance) is about 75. The free spectrum range (110 cm in optical length) is about 300 MHz. The coupling beam is rejected by the polarized beam splitter 4 (PB4) before reaching the detector.

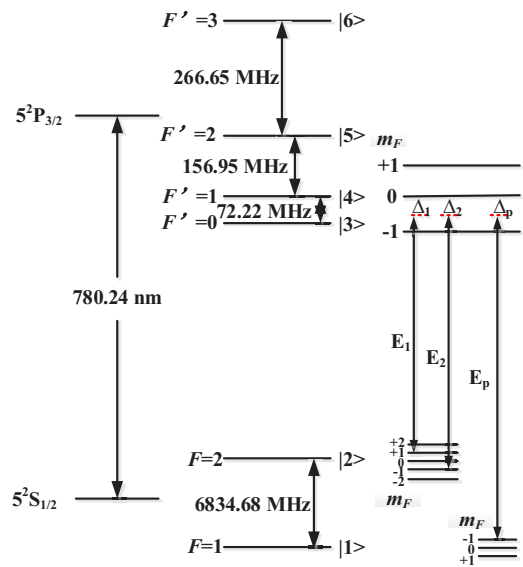

Fig. 1. (Color online) The relevant energy levels of $^{87}$Rb for our experiment.

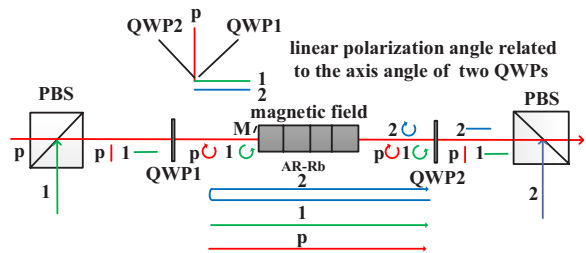

Fig. 2. (Color online) Detailed polarization states of three optical beams: p, probe beam $E_p$; 1, coupling beam $E_1$; 2, coupling beam $E_2$; PBS, polarizing cubic beam splitter; QWP1, QWP2, quarter-wave plates; M, mini-reflective mirror reflecting coupling beam $E_2$; AR-Rb, anti-reflection coated Rb vapor cell.

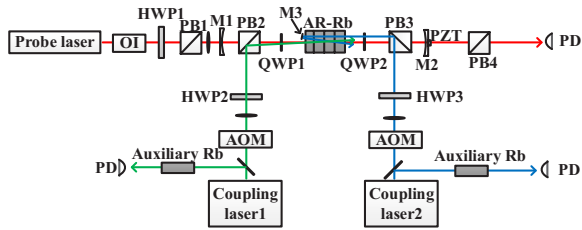

Fig. 3. (Color online) Schematic diagram of the experimental setup: PB1-PB4, polarizing cubic beam splitters; HWP1-HWP3, half-wave plates, QWP1, QWP2, quarter-wave plates; AOM, acoustic optical modulator; PD, photo-diode detector; OI, optical isolator; M1, M2, cavity mirrors; M3, mini-reflective mirror; AR-Rb, anti-reflection coated Rb vapor cell.

First, we observe the free space EIT spectrum as the mirrors M1 and M2 are removed. The probe beam frequency is scanned through the transition F=1 to F'=1 while employing a static frequency of coupling beam $E_1$ and coupling beam $E_2$. The Rb vapor cell is kept at room temperature. We apply a magnetic of 107.14 G in our experiment to make atomic levels split into Zeeman sublevels and the frequency separation between the sublevels of F=2 is 75 MHz. As is discussed in [17, 19, 20], in order to get a narrow EIT window, the intensity and detuning of the two coupling fields should be different. In our experiment, the AOMs are modulated to make the two coupling fields resonant to level F'=1 with a frequency separation of 170 MHz. Thus, taking the transition rule of Zeeman sublevels and the polarization states of optical beams into consideration, the two coupling fields have a detuning of $\Delta_1 = -10$ MHz and $\Delta_2 = +10$ MHz (detuning to level F'=1, $m_F = 0$, as is shown in Fig. 4). The probe laser is approximately $E_p = 48\mu$W and the two coupling fields are $E_1 = 1.63$ mW, $E_2 = 0.49$ mW respectively. Therefore, EITs occur when $\Delta_p = -10$ MHz and $\Delta_p = +10$ MHz. Figure 5 shows the probe field transmission versus the detuning of the probe field. From this figure, we can see that the EIT at $\Delta_p = +10$ MHz is much narrower than the other one. According to the previous study result[23, 24], the rate of the narrowed cavity linewidth $\Delta\omega$ to that of the empty cavity $C$ reads

$$\frac{\Delta\omega}{C} \propto \frac{1}{1 + \omega_0(l/2L)\eta}, \quad (1)$$

where $l$ is length of the vapor cell, $L$ is the cavity length, $\omega_0$ is the resonance frequency which are the order of $10^{14}$ Hz for the Rb atoms. Then, it is easy to see that the narrowed cavity linewidth is in roughly reverse proportion to the dispersion $\eta$. Both our previous theoretical analysis and present experimental result show that the interacting dark states lead to a narrower transparency window and the larger dispersion exists in the narrower EIT window of the double-dark system. Therefore, the ultranarrow cavity transmission spectrum (CTS) could be expected via this EIT window.

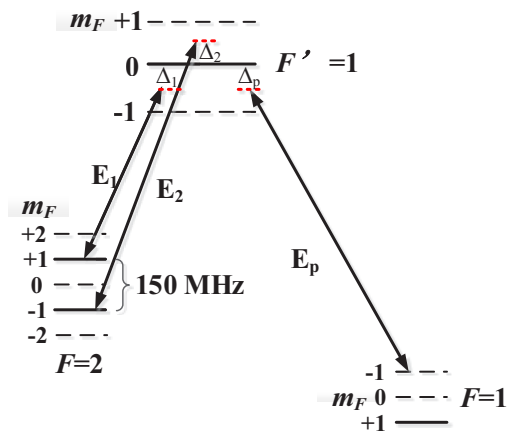

Fig. 4. (Color online) Tripod-type EIT system for our experiment: $E_p$, probe beam; $E_1, E_2$, coupling beams.

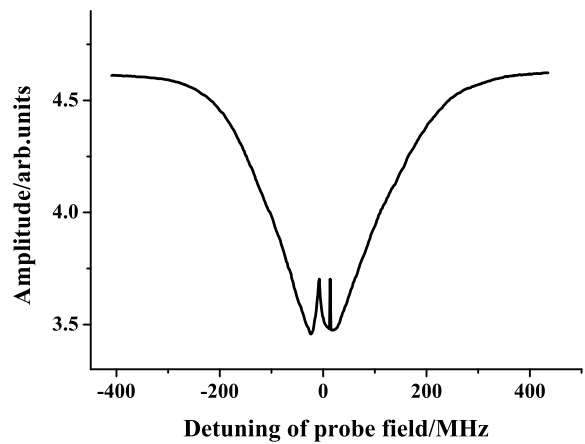

Fig. 5. Two transparency windows induced by the double-dark resonances EIT system: the right window ($\Delta_p=\Delta_2$) is much narrower than the left one ($\Delta_p=\Delta_1$) as $E_2$=0.49 mW is weaker than $E_1$=1.63mW.

Then, we put the mirrors M1 and M2 back to form an FP cavity, as Fig. 3 shows. To match the resonant frequency $\omega_r$ of the cavity plus double-dark resonances medium to the probe field frequency $\omega_p$ at $\Delta_p = +10$ MHz, the cavity length is adjusted by tuning the driving voltage of the PZT. When the condition ($\omega_r = \omega_{p,\Delta_p=+10\text{MHz}}$) is met, the transmission peak is high and narrow because of the steep dispersion led by the interacting dark resonances in the tripod-type EIT system. Figure 6 shows a comparison of the CTS linewidth for probe frequency well off the absorption line and the CTS linewidth when the intracavity resonances are formed. With double-dark resonances, the CTS linewidth is 550 KHz, which is 7 times narrower than the CTS linewidth with intracavity loss but the probe field off resonance. In the similar experimental condition, we also narrow the CTS linewidth using the single-dark resonance. The Rb cell is kept at room temperature without magnetic field. The coupling beam $E_1$ is removed and the coupling beam $E_2$ is resonant to level F'=1 with power of $E_2 = 0.49$ mW. As we adjust the cavity length to match the resonant frequency $\omega_r$ of the cavity plus single-dark resonance medium to the probe field frequency at $\Delta_p = 0$ MHz, the CTS linewidth is 1.55 MHz. It is about 3 times wider than the one obtained in the double-dark resonances case.

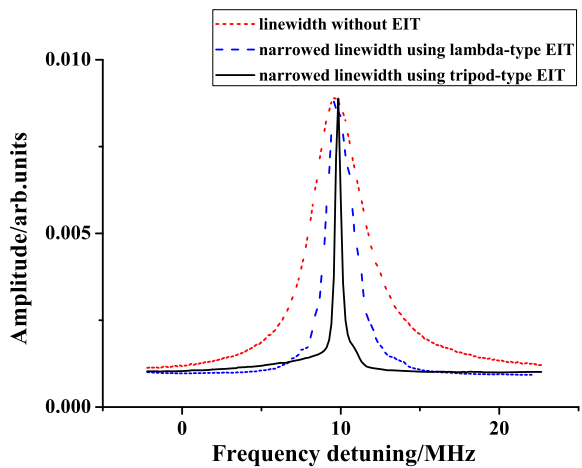

Fig. 6. (Color online) Intensity of cavity output versus probe field detuning, showing CTS linewidth narrowing in different conditions.

It is important to indicate that the measured CTS linewidth is not equal to the real cavity linewidth. For the input laser profile $I_{in}(\omega,\omega_0)$ with center frequency $\omega_0$, the cavity transmission intensity at $\omega_0$ should be

$$I_{out}(\omega_0) = \int_{-\infty}^{\infty} I_{in}(\omega,\omega_0) \cdot T(\omega) d\omega, \quad (2)$$

where $T(\omega)$ is the cavity transmission function and $\omega$ is laser frequency. As the functions $I_{in}(\omega)$ and $T(\omega)$ usually approximate to the Lorenz profile, the transmission intensity

$$\begin{aligned} I_{out}(\omega_0) &= \int_{-\infty}^{\infty} I_{in}(\omega,\omega_0) \cdot T(\omega) d\omega \\ &= \int_{-\infty}^{\infty} I_{in}(\omega - \omega_0) \cdot T(\omega) d\omega \\ &= \int_{-\infty}^{\infty} I_{in}(\omega_0 - \omega) \cdot T(\omega) d\omega \\ &= I_{in}(\omega_0) * T(\omega_0), \end{aligned} \quad (3)$$

where $*$ means the convolution operation. Because of the features of Lorenz function in the convolution operation, the width of the profile $I_{out}(\omega_0)$ equals the width of the profile $I_{in}(\omega_0)$ plus the width of the profile $T(\omega_0)$. As Fig. 6 is plotted with scanning the center frequency of probe beam, so $\omega_0$ corresponds to the abscissa of Fig. 6 and the width of profile $I_{out}(\omega_0)$ corresponds to the CTS linewidth in Fig. 6. The width of the profile $T(\omega_0)$ equals the cavity linewidth. Thus, the CTS linewidth obtained in our experiment is about the sum of cavity linewidth and probe laser linewidth. In order to extract the real cavity linewidth from the CTS linewidth, we measure the probe laser linewidth using the beat frequency method. For the measured probe laser linewidth of 300 KHz, the cavity linewidth with double-dark resonances obtained in our experiment is only about 250 KHz, which is 5 times narrowed than the single-dark resonance case.

On the other hand, the narrowed cavity linewidth would become broader for the single-dark resonance case when the coupling field is detuned. As our theoretical analysis in [28], the ultranarrow cavity linewidth would keep in a wide frequency range in the double-dark resonances system. In order to make a detailed research on how the frequency detuning affects the cavity linewidth, we measure the narrowed cavity linewidth under different detuning in both the single-dark and double-dark cases. In the experiment, we record the cavity linewidth as the detuning of the coupling beam $E_2$ changes from $\Delta_2 = -45$ MHz to $\Delta_2 = +45$ MHz. The detuning of the coupling beam $E_1$ is kept at $\Delta_1 = -50$ MHz in the double-dark case to make the two EITs separated to each other. The experimental result is shown in Fig. 7. As the detuning is large ($\Delta_2 < -30$ MHz or $\Delta_2 > 30$ MHz), the obtained narrowed cavity linewidth in the double-dark case approximates to the single-dark case. As the detuning is small (-30 MHz$< \Delta_2 < +30$ MHz), we can see that the ultranarrow linewidth keeps at about 250 KHz with deviations less than 15 KHz in the 60 MHz detuning range in the double-dark resonances case, while the linewidth is broadened from 1250 KHz to 1390 KHz in the single-dark resonance case. This effect may have potential applications in tunable high-resolution spectroscopy and laser frequency stabilization.

Finally, there are several experimental details need to be considered in order to obtain a good cavity linewidth narrowing effect via double-dark resonances with the tripod-type Zeeman splitting sublevels. The most im-portant one is the magnetic amplitude. As is well known, when the probe and coupling laser fields are applied to Zeeman splitting sublevels, each transitions which meet the transition rule of Zeeman sublevels are possible. Taking the coupling beam $E_1$ for example, it may drive three transitions which meet the condition of $\Delta m_F = -1$ ($m_F = +2 \rightarrow m_F = +1$, $m_F = +1 \rightarrow m_F = 0$, $m_F = 0 \rightarrow m_F = -1$), as is shown in Fig. 8. When a weak magnetic is applied, the amount of Zeeman levels shift is small and then the above three transition frequencies are approximately equal. As a result, these three transitions are all strongly coupled with the coupling beam $E_1$ and several individual EIT subsystems are formed. It would disturb the tripod-type EIT observation. In our experiment, applying a magnitude field with amplitude of about 100 G, there is a differentiation of more than 25 MHz in each transition frequency. As a result, the disturbance of existed additional Zeeman splitting sublevels is not significant. In addition, the current experimentally achieved cavity-linewidth narrowing effect is still limited by some other factors, including the intracavity losses (due to the surfaces of the vapor cell, the PBs and the QWPs), the reduced transparency by the optical pumping effect of the strong coupling field and the mechanical vibration of the optical cavity. With further technical improvements, the narrowed cavity linewitdth which is one order of magnitude narrower than that of single-dark system can be expected, as the theoretical predicated in [28].

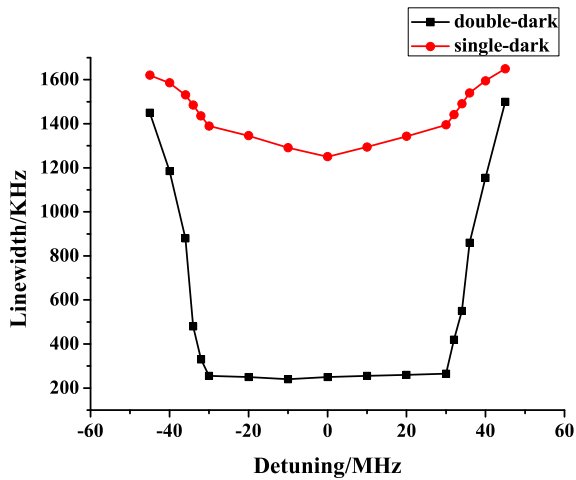

Fig. 7. (Color online) Cavity linewidth versus the detuning of coupling field.

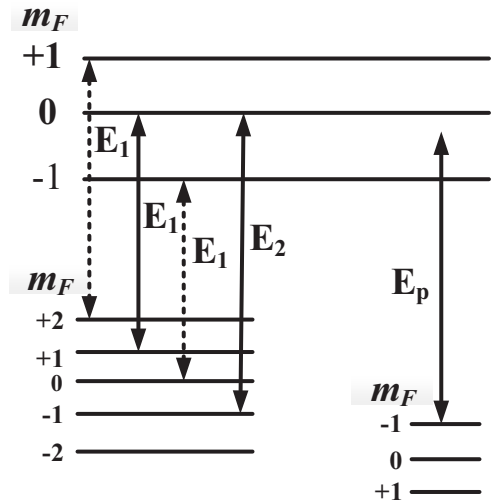

Fig. 8. Zeeman sublevels coupled by circularly polarized light.

## 4. Conclusion

In conclusion, we have experimentally demonstrated cavity linewidth narrowing by the interacting double-dark resonances in a tripod-type EIT system. Applying a magnitude field with amplitude of about 100 G, a tripod-type configuration is formed using $^{87}$Rb atoms Zeeman splitting sublevels. With the steep dispersion in the narrower tripod-type EIT windows, the cavity linewidth is narrowed to about 250 KHz. It is 5 times narrower than the single-dark resonance case at room temperature. Furthermore, the ultranarrow linewidth keeps at about 250 KHz with deviations less than 15 KHz in the 60 MHz detuning range, which have potential applications in tunable high-resolution spectroscopy and laser frequency stabilization.

This work was supported by the National Natural Science Foundation of China (Grant Nos. 11274112, 61108028 and 61178031) and Shanghai Rising-Star Program of Grant No. 11QA1407400.